\documentclass[10pt,preprint]{aastex63}

\graphicspath{{plots/}}

\usepackage{mathrsfs, amsmath}
\usepackage{dblfloatfix}
\usepackage{csquotes}
\usepackage{dblfloatfix}
\usepackage{fixltx2e}
\usepackage{url}
\usepackage{natbib}

\usepackage{graphicx}
\usepackage{enumitem}

\submitjournal{PASP}
\received{Dec. 20, 2022}
\revised{Jan. 23, 2023}
\accepted{Jan. 25, 2025}

\begin{document}
\title{High Speed Focal Plane Wavefront Sensing with an Optical Chopper}
\correspondingauthor{Benjamin L. Gerard}
\email{gerard3@llnl.gov}
\author[0000-0003-3978-9195]{Benjamin L. Gerard}
\affiliation{Lawrence Livermore National Laboratory}
\affiliation{University of California Santa Cruz}

\author{Daren Dillon}
\affiliation{University of California Santa Cruz}

\author{Sylvain Cetre}
\affiliation{Durham University}
\affiliation{Wakea Consulting}

\author[0000-0003-0054-2953]{Rebecca Jensen-Clem}
\affiliation{University of California Santa Cruz}
\begin{abstract}
Focal plane wavefront sensing and control is a critical approach to reducing non-common path errors between the a conventional astronomical adaptive optics (AO) wavefront sensor (WFS) detector and science camera. However, in addition to mitigating non-common path errors, recent focal plane wavefront sensing techniques have been developed to operate at speeds fast enough to enable ``multi-WFS'' AO, where residual atmospheric errors are further corrected by a focal plane WFS. Although a number of such techniques have been recently developed for coronagraphic imaging, here we present one designed for non-coronagraphic imaging. Utilizing conventional AO system components, this concept additionaly requires (1) a detector imaging the focal plane of the WFS light source and (2) a pupil plane optical chopper device that is non-common path to the first WFS and is synchronized to the focal plane imager readout. These minimal hardware requirements enable the temporal amplitude modulation to resolve the sine ambiguity of even wavefront modes for both low, mid, and high wavefront spatial frequencies. Similar capabilities have been demonstrated with classical phase diversity by defocusing the detector, but such techniques are incompatible with simultaneous science observations.  This optical chopping techniqe, however, enables science imaging at up to a 50\% duty cycle. We present both simulations and laboratory validation of this concept on SEAL, the Santa Cruz Extreme AO Laboratory testbed.
\end{abstract}
\section{Introduction}
\label{sec: intro}
Astronomical adaptive optics (AO) has enabled ground-breaking discoveries over the past three decades, from the Nobel prize-winning imaging and dynamical analysis of the Galactic center \citep{ghez} to the first images of planets around other stars on Solar System scales \citep{marois}. However, advancements over the years has illuminated a new limitation to further improving performance of AO systems: non-common path aberrations (NCPAs) between an AO wavefront sensor (WFS) and AO-fed science detector. These NCPAs are fundamentally not measurable with a standard AO WFS and therefore not correctable (spatially or temporally) by an AO deformable mirror (DM) without additional wavefront sensing methods. 

NCPAs are among the dominant terms in error budgets of current AO systems \citep[e.g.,][]{poyneer}, and as such there has been significant recent development of NCPA correction techniques for AO systems \citep[e.g.,][]{bottom16,potier22}. Considering the amplitude, $A$, and phase, $\phi$, of the electromagnetic field from a star at infinity placed on the DM of an AO system (i.e., conjugated to the telescope pupil), a downstream focal plane point spread function (PSF) image is described by
\begin{equation}
    \mathrm{PSF}=|\mathcal{F}\{A\; e^{i\; \phi}\}|^2, \nonumber
\end{equation}
where $\mathcal{F}$ is a Fourier transform operator, $|\; |^2$ represents square modulus of the focal plane electric field and $i$ is $\sqrt{-1}$. A PSF at perfect focus is inherently degenerate to certain modes of $\phi$, including the sine of symmetric Zernike modes (e.g., focus, spherical aberration, etc.) and the relative phase of Fourier modes (e.g., a 3 cycle/pupil sine vs. cosine), and therefore additional techniques are needed to sense the wavefront from a science image and then update DM commands and/or AO WFS offsets to accordingly compensate for NCPAs.

NCPA measurement and correction techniques, also known as focal plane wavefront sensing and control, generally fall into one of two categories to resolve the wavefront measurement degeneracy with a PSF: temporal or spatial modulation. Temporal modulation involves recording a series of two or more images with the science camera where a known probe is applied between images, resolving the wavefront ambiguity with ``temporal diversity'' \citep[e.g.,][]{phase_diversity,speckle_nulling}. Conversely, techniques with spatial modulation can in principle use a single focal image for wavefront reconstruction, leveraging custom hardware and/or software to resolve the wavefront measurement degeneracy, e.g., with fringes \citep{baudoz06} or calibrated symmetry \citep{miller17} in the focal plane image. 

In general, temporal modulation-based focal plane wavefront sensing techniques use iterative algorithms designed for stable, space telescope-like environments. In this paper, we introduce a temporal modulation-based focal plane WFS designed for correction of residual AO turbulence down to millisecond timescales using an optical chopper device. We note that this technique was originally introduced in \cite{gerard22spie}, but is expanded and presented more verbosely here. In \S\ref{sec: concept_and_sim} we present the concept of pupil chopping for focal plane wavefront sensing and present simulations. We then present validating laboratory results using the Santa Cruz Extreme AO Laboratory (SEAL) testbed in \S\ref{sec: lab}, provide further discussion in \S\ref{sec: discussion}, and then conclude in \S\ref{sec: conclusion}. Unless otherwise noted, simulations in this paper assume an operating wavelength of $\lambda=1.6\mu$m, a 256$\times$256 pixel image size, and beam ratio (number of pixels per $\lambda/D$) of 3.
\section{Concept and Simulations}
\label{sec: concept_and_sim}
\subsection{Concept}
\label{sec: concept}
\begin{figure}
\includegraphics[width=0.5\textwidth]{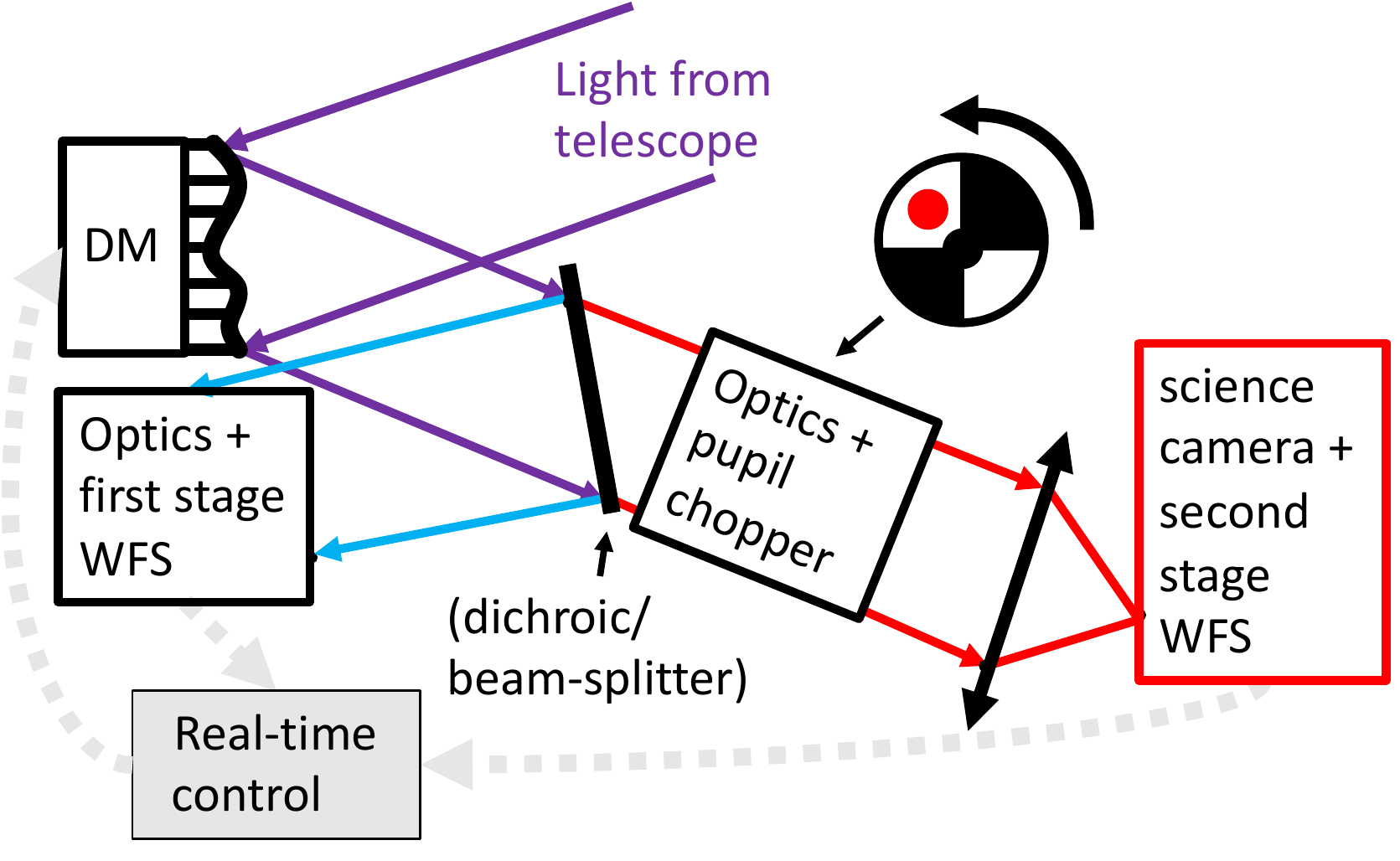}
\includegraphics[width=0.5\textwidth]{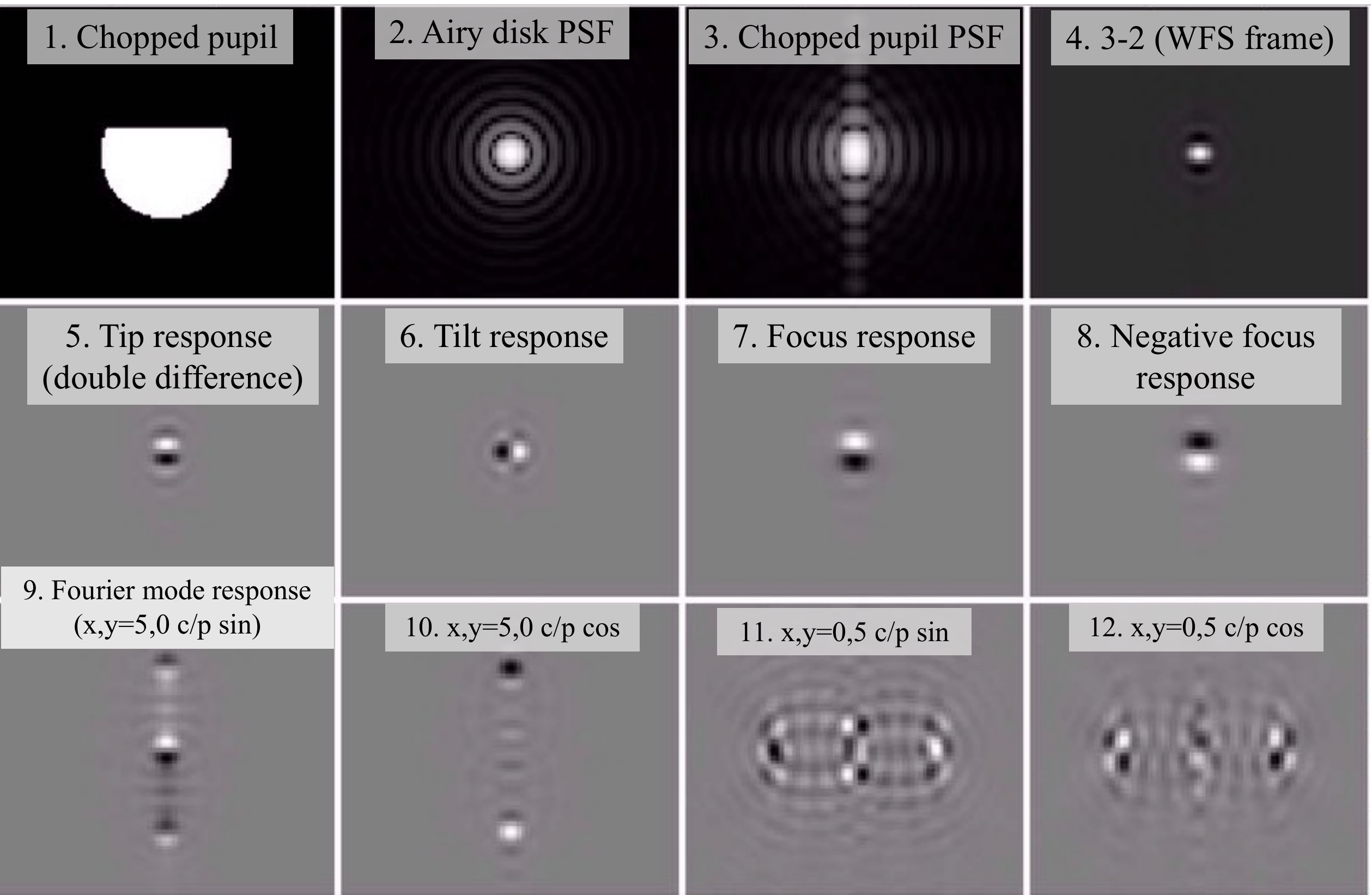}
\hspace*{150pt} (a) \hspace{200pt} (b) 
\caption{Illustration of the pupil chopping concept, both from a geometric (a) and Fourier optics (b) perspective. In (b), all focal plane images are shown with the same 10$\times$10 $\lambda/D$ field of view. Panels 2-3 are shown on a log scale; all other panels are shown on a linear scale. Zernike and Fourier modes in panels 5-12 all use 1 nm amplitudes. Panels 7-12 indicate that pupil chopping for focal plane wavefront sensing can resolve the sign ambiguity of focus and the relative phase of Fourier modes, which is not possible with a single PSF image.}
\label{fig: concept}
\end{figure}

Figure \ref{fig: concept} illustrates the concept of focal plane wavefront sensing with a pupil plane optical chopper.
Two focal plane images are required to enable a WFS measurement: one with the chopper blade partially blocking the pupil, and one with the pupil unblocked; the latter can be used for science while the former is limited to wavefront sensing. A ``reference'' WFS frame is saved before on-sky operations and then used analogously to the concept of on-sky reference slopes for other pupil plane WFSs. Such reference slopes enable a real-time WFS frame to measure the wavefront down to the flatness level of the pre-calibrated reference image. In equation form, the on-sky WFS frame, $w$, at an instance in time is given by
\begin{equation}\label{eq: chop_im_onsky}
\begin{split}
w &= \text{chop}\{ \phi_\text{on-sky} \} - \text{chop}\{ \phi_\text{ref} \},\text{ with} \\
\text{chop}\{\phi\} &\equiv \left( |\mathcal{F}\{A_\text{chop} e^{i \phi}\}|^2 - |\mathcal{F}\{A_\text{norm} e^{i \phi}\}|^2 \right)/\Sigma\{ |\mathcal{F}\{A_\text{norm} e^{i \phi}\}|^2\}, \nonumber
\end{split}
\end{equation}
where $A_\text{norm}$ is an unobscured pupil wavefront amplitude (i.e., an open slot of the chopper wheel), $A_\text{chop}$ represents a chopped pupil wavefront amplitude, $\phi_\text{on-sky}$ is the on-sky pupil wavefront phase, $\phi_\text{ref}$ is the reference pupil wavefront present during the above mentioned daytime calibration, and $\Sigma\{ \}$ represents an operator for summing pixel values within a given image. To be clear, $\phi_\text{ref}$ cannot be generated or reconstructed with this technique, and must be obtained by some other non-linear iterative wavefront sensing and/or phase retrieval technique to flatten the absolute wavefront at the science detector plane (e.g., with phase diversity from \citealt{lamb17} and/or the asymmetric pupil mask from \citealt{martinache13}; see \S\ref{sec: discussion}). However, the advantage of this double difference approach is that it enables a linear wavefront reconstruction from $w$ space to DM command space (i.e., optimal for real-time wavefront control of AO residuals), which we will describe and illustrate in the next section.

Due to the implementation of a continuously spinning blade, the duty cycle for a chopped or un-chopped image is limited to a maximum of $\frac{1}{2}(1-\frac{d_\text{beam}}{d_\text{blade}})$, where $d_\text{beam}$ is the beam diameter and $d_\text{blade}$ is the chopper blade width (assuming equal spacing between obscuring and unobscuring slots of the blade; so if $d_\text{blade}=d_\text{beam}$, there is no time to keep the beam un-chopped and the duty cycle is 0, whereas the maximum duty cycle with a very tiny pupil and/or large blade, $\frac{d_\text{beam}}{d_\text{blade}}\approx0$, is 0.5). To clarify, for this technique $d_\text{blade} \ge d_\text{beam}$ (i.e., a blade size smaller than the beam size would prevent either frame from seeing a unobscured pupil). A phase delay is also needed to acquire chopped and un-chopped pupils in consecutive focal plane images, without which in every other frame the pupil would be either completely blocked and then un-blocked or chopped by a fraction, $f$, on one side and then symmetrically chopped by 1-$f$ on the other side in the next frame.

By being a focal plane wavefront sensor, this method benefits from natural wavefront spatial filtering properties (i.e., with the focal plane image representing the Fourier plane of the pupil plane wavefront). This configuration enables binary masks to be placed on the focal plane image WFS that act as a natural anti-aliasing filter and minimize non-linear cross talk between modes. Using the focal plane image and such binary masks for wavefront sensing with this technique also distinguishes this technique from the differential optical transfer function method \citep{codona13}. This method is also complementary to pupil amplitude diversity-based absolute phase retrieval methods, such as the asymmetric pupil Fourier WFS \citep{martinache13}. Our pupil chopping technique enables a linear-least squares reconstruction (as we will show next in \S\ref{sec: lowfs_recon_sim}), optimal for high speed wavefront control of AO residuals but not able to flatten the absolute phase below a pre-calibrated best flat (e.g., due to evolving on-sky quasi-static aberrations), whereas phase retrieval methods can use the chopped pupil PSF image to recover the absolute phase to track such quasi-static effects, but typically requiring non-linear iterative algorithms that cannot be run at high speed. See \S\ref{sec: discussion} for a further discussion about combining these two complementary approaches.
\subsection{Low Order Wavefront Reconstruction and Control}
\label{sec: lowfs_recon_sim}
With the setup described in \S\ref{sec: concept}, we enable a linear matrix vector multiply (MVM) reconstructor for real-time wavefront control as described in Appendix \ref{sec: mvm}. Figure \ref{fig: lowfs_sim_recon} shows the result of open loop MVM-based modal coefficients for 13 Zernike modes.
\begin{figure}[!h]
\centering
\includegraphics[width=0.5\textwidth]{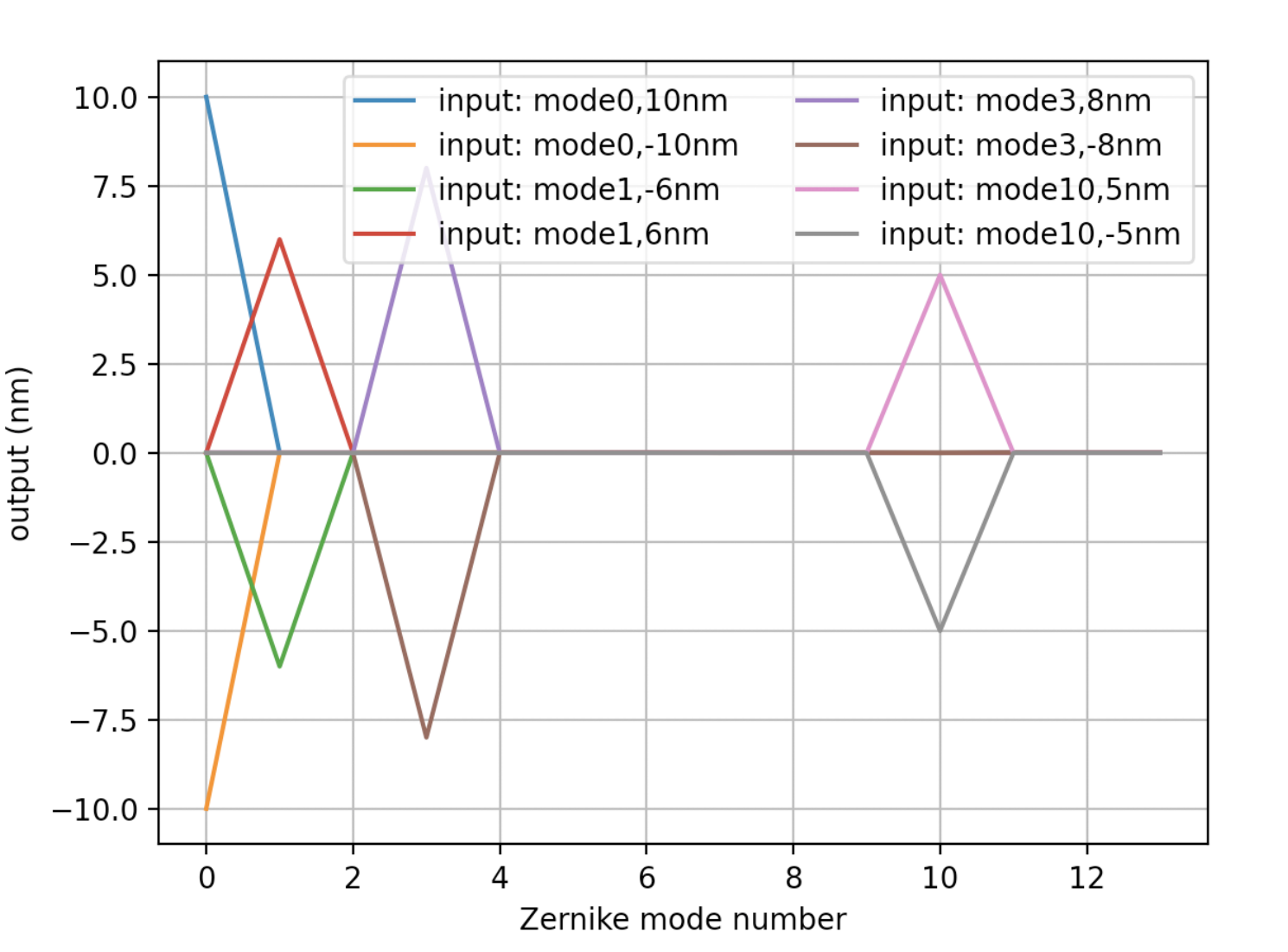}
\caption{Simulated open-loop reconstruction of modal coefficients for different Zernike modes and amplitudes (mode 0 in this case refers to tip). The peak-to-valley IM amplitude is 1 nm for all modes, mapping modal coefficients to reconstructed wavefront in nm. }
\label{fig: lowfs_sim_recon}
\end{figure}
Fig. \ref{fig: lowfs_sim_recon} clearly shows the potential of this focal plane wavefront sensing technique as a linear WFS for input wavefront phases with wavefront errors less than 1 radian rms (i.e., the standard regime in which a Taylor expansion of the PSF is linearly related to the pupil wavefront phase).
\subsection{Limitations to High Order Wavefront Reconstruction}
\label{sec: ho_sim}
\begin{figure}[!h]
\centering
\includegraphics[width=0.9\textwidth]{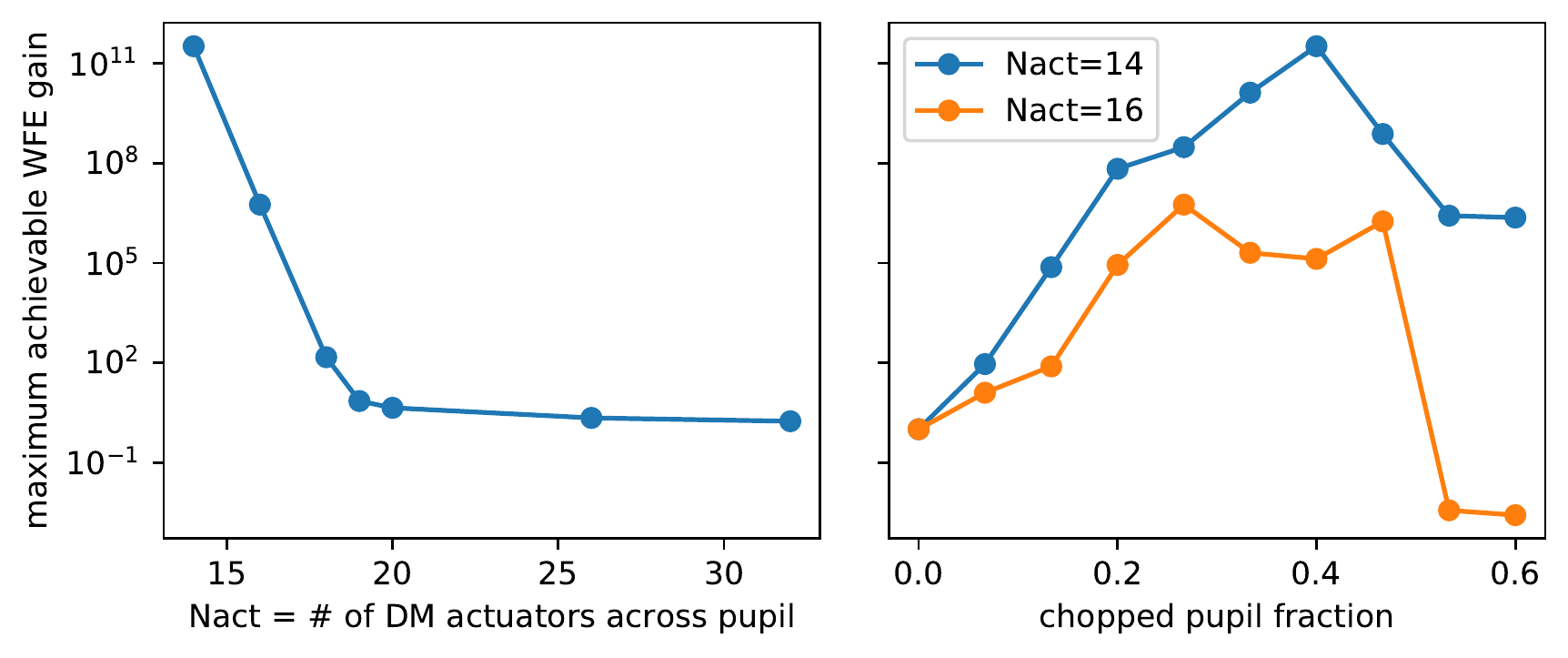}
\caption{Grid search simulations, optimizing all adjustable parameters in the wavefront reconstruction code as a function of DM actuator count (left) and chopped pupil fraction for two different actuator counts (right). The y-axis for both panels shows the ratio of input (100 nm rms) to convergent output wavefront errors due to non-linearities (i.e. without aliasing or photon noise) after 12 iterations, medianed over 100 random realizations for each data point shown. Interaction matrix amplitude, SVD cutoff, and the radius for a binary mask to isolate sine spots from Fourier modes are optimized for all data points in both panels, with chopped pupil fraction additionally optimized for the left panel.}
\label{fig: ho_limit_sim}
\end{figure}

Using this pupil chopping technique, we carried out a detailed analysis of achievable residual wavefront error on input extreme AO residual turbulence (i.e., a static phase screen normalized to 100 nm rms from 0 to 10 c/p with a -2 power law, similar to \citealt{poyneer}, with results medianed over 100 different random realizations) as a function of DM actuator count, illustrated in Figure \ref{fig: ho_limit_sim}. 
The left panel of Fig. \ref{fig: ho_limit_sim} shows that pupil wavefront spatial frequencies greater than around 9 c/p do not converge to a deeper WFE than the input AO residual level, while spatial frequencies less than this limit in principle can gain in convergent vs. input WFE by many orders of magnitude. Intuitively, this 9 c/p wavefront measurement limit is because light compared with an un-modulated Airy disk is only modulated in a chopped frame out to about 9 $\lambda/D$ with the current chopper blade concept. We will propose possible solutions to this limit and discuss this further in \S\ref{sec: discussion}. The right panel of Fig. \ref{fig: ho_limit_sim} further illustrates performance dependence on what fraction of the pupil is chopped, showing that $\sim$30-50\% enables sufficient gains, which will justify our later laboratory implementation in \S\ref{sec: lab}.
\section{Laboratory Testing}
\label{sec: lab}
\subsection{Setup}
\label{sec: setup}
Our laboratory setup, mimicking Fig. \ref{fig: concept}a, uses SEAL---the Santa Cruz Extreme AO Laboratory (\citealt{gerard22}, \citealt{jensen_clem21}). We use a Thorlabs KLS635 laser (set at 0.15 mW unless otherwise noted),  ALPAO 97 actuator DM (which also serves as the system pupil stop, Andor Zyla 5.5 sCMOS detector, Thorlabs WFS-20 SHWFS (for this paper used only to flatten the ALPAO DM), a Thorlabs MC2000B optical chopper with blade MC1F10, and a Stanford Instruments DG535 controller and an in-house electrical doubler so that the Andor detector readout is synchronized with the optical chopper (with the chopper as the leader and the Andor camera as the follower\footnote{Although the AO field has historically adopted ``master and slave'' terminology from the computer science field to refer to the chopper and camera in this configuration, respectively, it is now clear that such discriminatory terminology should no longer be used. Following the computer science field, instead we will refer to such configuration as ``leader and follower'' throughout this paper, encouraging other members of the AO field to do the same.}) to produce a chopped and un-chopped pupil every other frame, respectively. As in \cite{gerard22}, ALPAO DM units are converted to WFE units via a single scalar multiplication, calibrated via open tip/tilt measurements and corresponding PSF images, but for which higher order conversions become increasingly wrong due to the decreasing stroke limits of the DM as a function of mode order.

Unless otherwise noted, our standard laboratory configuration for chopper WFS image acquisition is with the chopper blade running at 100 Hz, using the doubler so the Andor camera follower runs at 200 Hz with a 0.1 ms exposure per frame (i.e., a 2 \% duty cycle), and also implementing a 2.5ms phase delay for the doubled chopper signal with the Stanford controller. As in \cite{gerard22}, we also implement a serial 20 ms pause in between acquiring a chopper pair of images and sending any DM commands (both for calibration and real-time control software) due to additional latency from our Python interface to hardware components. Two random chopper imager pair acquisitions are not necessarily in the same order (i.e., one could have the chopped frame first while the other could have the un-chopped frame first), but for consistency we order every chopper sequence with the chopped frame first, as in Equation \ref{eq: chop_im_onsky}, determined as the frame with less cumulative flux.
A serial timing diagram outlining the procedure described in this section, which is used for real-time control, is shown in Table \ref{tab: timing}.
\begin{table}[!h]
    \centering
    \begin{tabular}{|c|c|c|c|c|c|c|c|c|c|}
        \hline
        \textbf{time (ms)} & 0 & 0.1 & 5 & 7.5 & 7.5 & 7.6 & 12.5 & 32.5 \\
        \hline
        \textbf{\shortstack{loop\\component}} & \multicolumn{2}{c|}{$|\mathcal{F}\{A_\text{chop} e^{i \phi_\text{on-sky}}\}|^2$} & \multicolumn{2}{c|}{phase delay} & \multicolumn{2}{c|}{$|\mathcal{F}\{A_\text{norm} e^{i \phi_\text{on-sky}}\}|^2$} & \multicolumn{2}{c|}{20 ms pause} \\
        \hline
        \textbf{\shortstack{loop\\timing}} & \shortstack{start\\camera\\integration} & \shortstack{stop\\camera\\integration} & \shortstack{finish\\camera\\duty\\cycle} & \shortstack{end\\delay} & \shortstack{start\\camera\\integration} & \shortstack{stop\\camera\\integration} & \shortstack{finish\\camera\\duty\\cycle} & \shortstack{apply\\DM\\ commands} \\
        \hline

    \end{tabular}
    \caption{Chopper timing diagram used for real-time control in \S\ref{sec: real_time_control}. }
    \label{tab: timing}
\end{table}
Due to the continuous sequence of images (camera integrations starting at 0, 7.5, 10, 17.5, 20, ... etc ms) being a non-integer multiple of the 32.5 ms total loop time shown in Table \ref{tab: timing}, a continuous real-time buffer is constantly acquiring chopped and un-chopped images and the real-time loop just grabs the most recent images from this buffer, meaning that effectively the 20 ms pause dominates the lag budget in our setup, is why subsequent real-time control results in \S\ref{sec: real_time_control} assume a $\sim$50 Hz loop rate.

Since the light source remains at a constant intensity, we also just use a single normalization value for all data rather than generate a new value for each chopper pair as equation \ref{eq: chop_im_onsky} suggests, but for changing light source intensities a moving average cumulative intensity should be considered to minimize noise propagation into this normalization term. The 0.15 mW setting for our testbed light source is set to place the PSF core at $\sim$2000 analog to digital units (ADUs, $\sim$half the Andor Zyla's 4196 ADU dynamic range), which from photon scaling relations \citep{bessell} for real-time purposes translates to a $\sim$90\% Strehl ratio from an extreme AO residual wavefront for 1 ms exposure, m$_H$=7 star, 3\% spectral bandpass centered at 1.6 $\mu$m (reasonable for a narrow band focal plane WFS setup to avoid blurring effects from PSF magnification with wavelength), and 50\% total sky-to-photoelectron throughput. Given that we can only control the 97 actuator ALPAO DM up to $\sim$5 cycles per pupil, only pixel values within a pre-defined 5$\lambda/D$ radius of the star are used for subsequent low order signal processing. The real-time differential image architecture of this technique lends itself to self-calibration, so dark subtraction or other ``cosmetic'' calibrations are not needed here. We found better performance when applying Zernike modes were non-illuminated actuators were set to the commands of their nearest neighbor rather than set to zero, and so we use the former hereafter in this paper unless noted otherwise.
\subsection{Low Order Linearity}
\label{sec: linearity}
\begin{figure}[!h]
    \centering
    \includegraphics[width=0.7\textwidth]{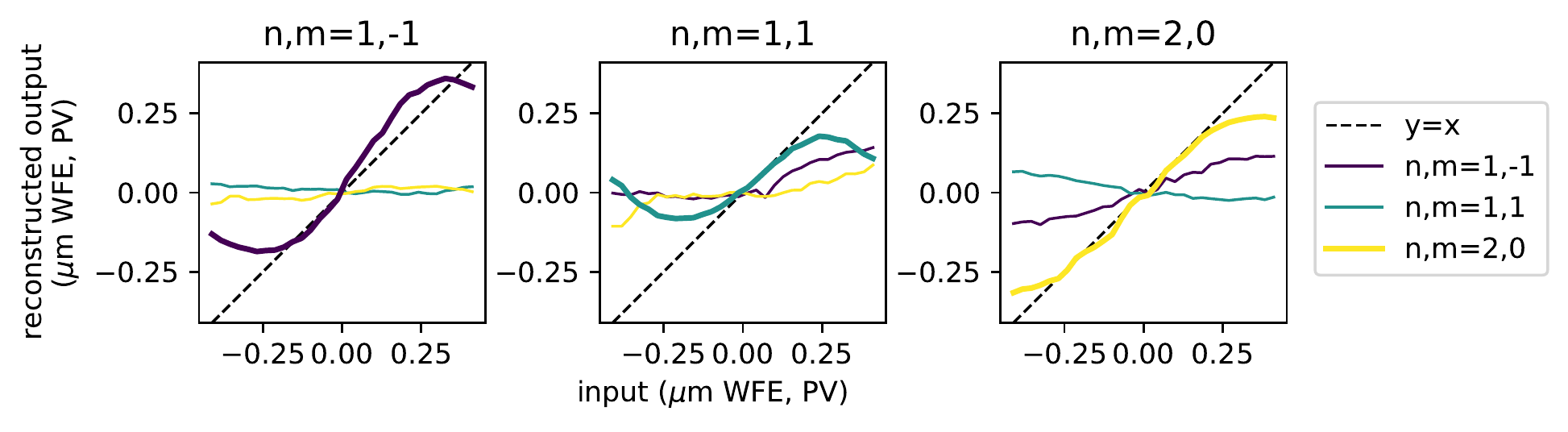}
    \\
    \hspace{-40pt} (a) Linearity for tip, tilt, and focus.
    \\[10pt]
    \includegraphics[width=0.8\textwidth]{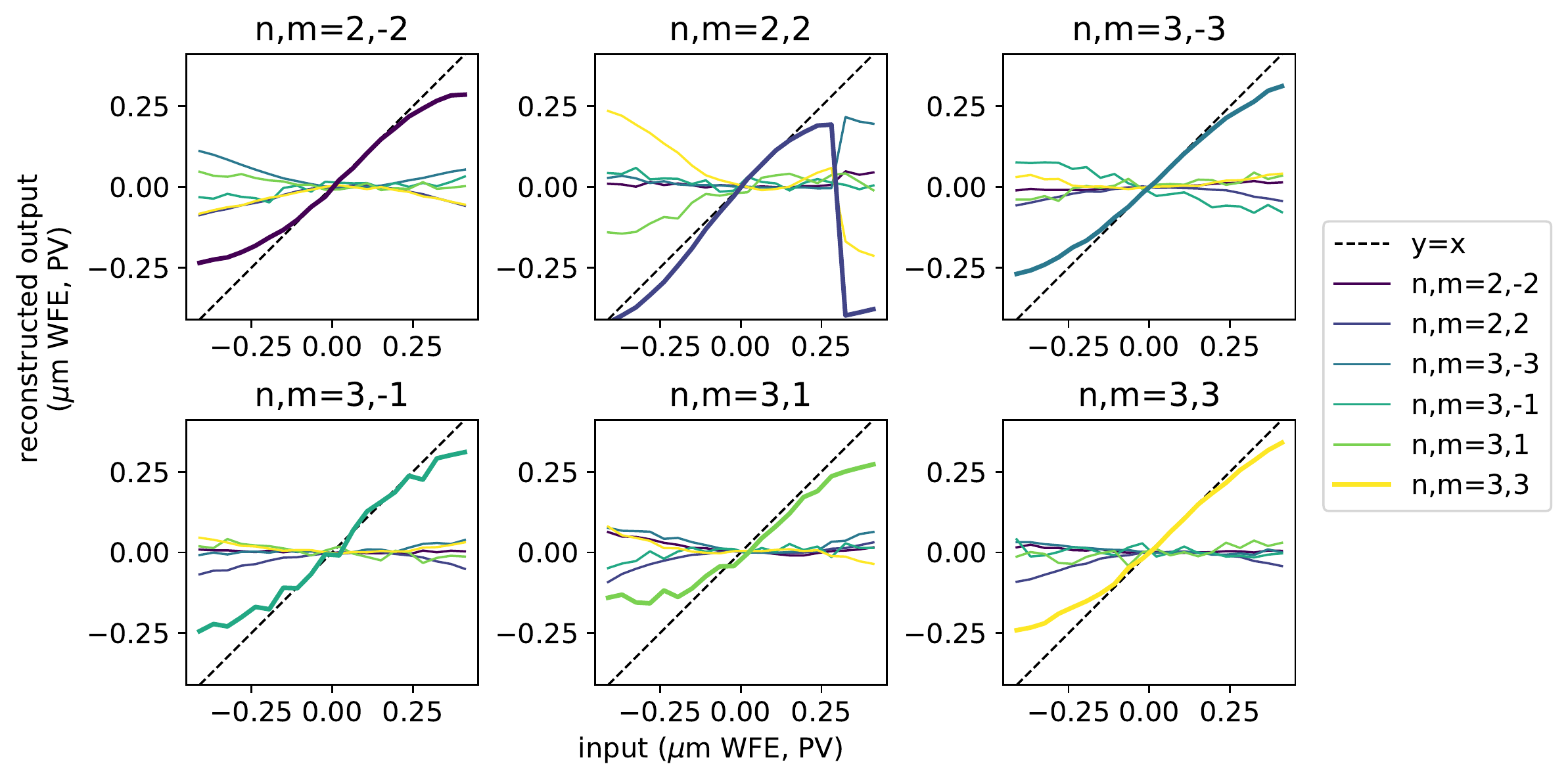}
    \\
    \hspace{-40pt} (b) Linearity for $n=2\rightarrow$3.
    \\[10pt]
    \includegraphics[width=0.8\textwidth]{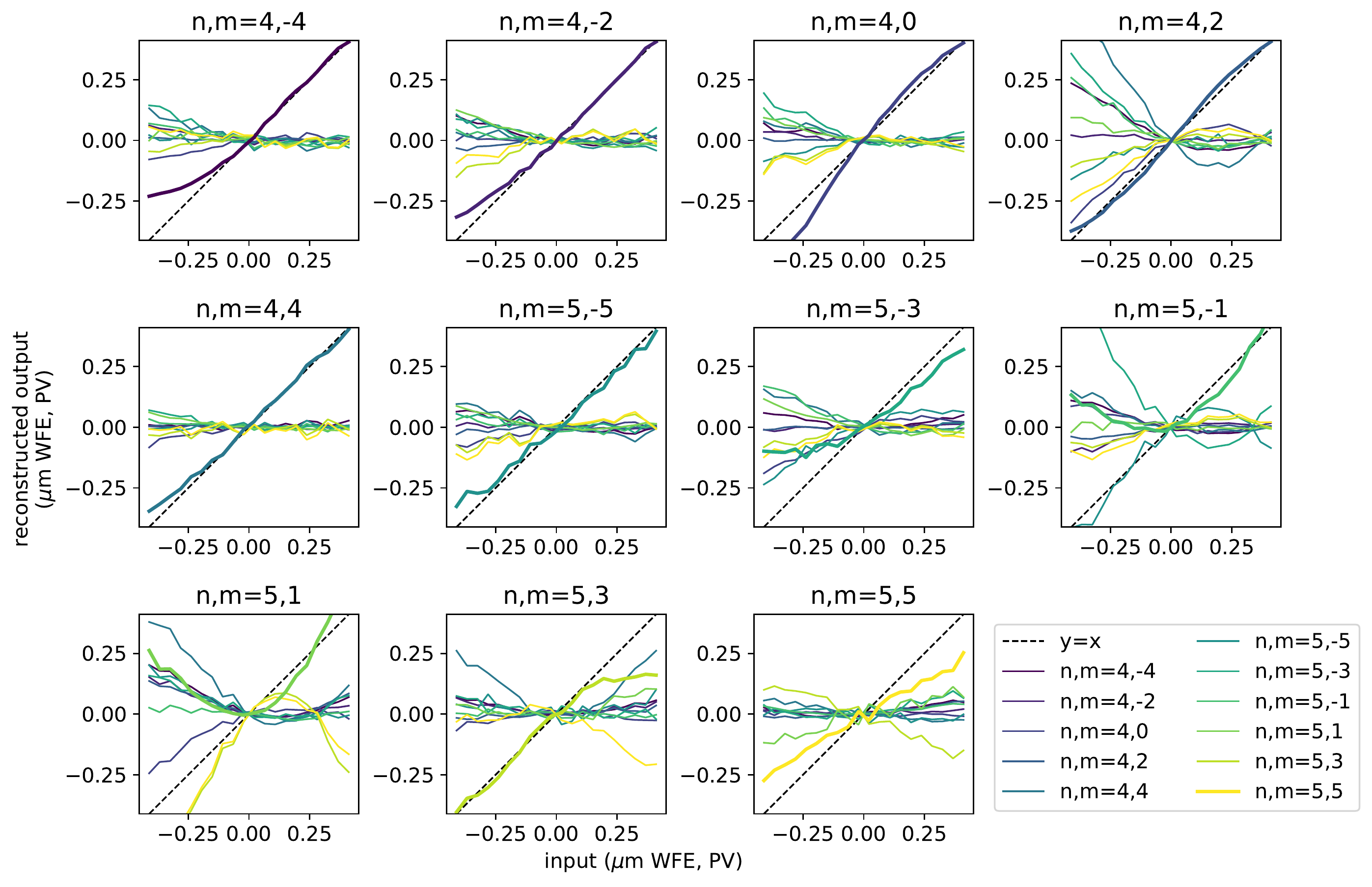}
    \\
    \hspace{-40pt} (c) Linearity for $n=4\rightarrow$5, informing our decision to not control the highly non-linear modes n,m=5,$\pm$1.
    \caption{Linearity curves for Zernike modes $n=1\rightarrow5$, separated into different modal groups (a-c), which are separately optimized by interaction matrix amplitudes and different command matrix SVD cutoffs.}
    \label{fig: linearity}
\end{figure}
After following the procedure outlined in \S\ref{sec: lowfs_recon_sim}, in Fig. \ref{fig: linearity} we generate reconstructed modal coefficients for low order Zernike modes, probing a range of input amplitudes for each mode in open loop and recording corresponding reconstructed output modal coefficients. In general, the cross terms in Fig. 5 are negligible\footnote{Ignoring n,m=5,$\pm$1 (which we use Fig. \ref{fig: linearity}c to motivate not controlling and accordingly note in the subfigure caption), in this context we use ``negligible'' to mean that for a given probed Zernike mode (1) it shows close-to-linear behavior for both positive and negative input amplitudes, and (2) cross terms do not reach amplitudes beyond the probed dominant mode over the linear range of (1). That being said, we acknowledge that decreased performance for both dominant mode and cross term non-linearities will result in decreased achievable closed-loop WFE and may require a modified calibration and/or control scheme, which are also discussed further in \S\ref{sec: real_time_control} and \S\ref{sec: discussion}.} when the given input mode is reconstructed within it's linear regime ($\lesssim$200 nm PV for each modal group, for tip/tilt corresponding to 5 mas for a 8m telescope at $\lambda=1.6\mu$m), which as expected is only feasible for AO residual WFEs. This also confirms that, as expected, a diffraction-limited AO residual wavefront is needed an an input  in order for this pupil chopping technique to enable closing the loop. Figure \ref{fig: linearity} also shows that linearities are computed separately in groups for (1) tip, tilt, and focus, astigmatism and coma, and (2) Zernike modes with n=4$\rightarrow$5. We found that the increased flexibility to use different SVD cutoffs for each group helped improve linearity and lower cross talk, motivating our choice to use this modal grouping. We also found that averaging 100 chopper frames per target image used for wavefront reconstruction to measure linearity gave significantly better stability than shorter sequences. The cause of such instability is described in the next section.
\subsection{Limitations to High Order Wavefront Reconstruction}
\label{sec: phase_jitter}
\begin{figure}[!h]
    \centering
    \includegraphics[width=0.5\textwidth]{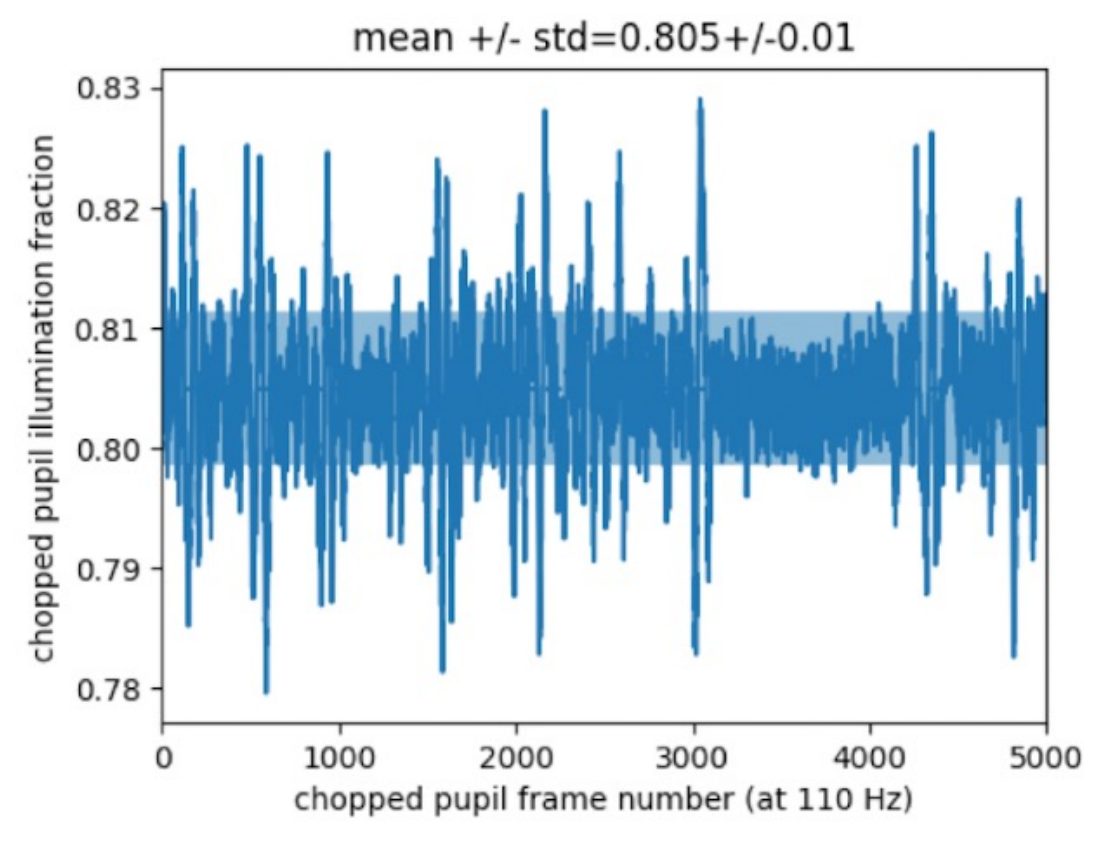}
    \caption{Chopper phase jitter analysis, showing the cumulative chopped pupil frame intensity as a fraction relative to the sequence-averaged cumulative un-chopped pupil frame intensity. The light source is saturating all illuminated pixels during this dataset, ensuring that variations shown here are due to chopper phase jitter, which are shown here to be present at the $\pm$1\% level, ultimately preventing high order wavefront reconstruction for spatial frequencies $\gtrsim2.5$ c/p.}
    \label{fig: phase_jitter}
\end{figure}

Unfortunately, we were not able to measure and control of Zernike modes with $n>5$ due to chopper blade phase jitter causing a time-varying chopped pupil fraction at the 1\% rms level, illustrated in Fig. \ref{fig: phase_jitter}.
This phase jitter is due to (1) chopper motor control variations and (2) imperfections of in the uniformity of the chopper blades. Our measurements in Fig. \ref{fig: phase_jitter} are consistent with the phase jitter specifications from Thorlabs, with no other available models able to reach smaller phase jitter levels. We found that $\gtrsim$100 nm amplitudes were needed for a given Fourier mode to detect sine spots at spatial frequencies $\gtrsim$3 c/p, which if used in the IM produces a quadratic linearity curve (i.e., creating a reconstructor that is unable to tell positive from negative input amplitudes). From simulations, $\lesssim$2 nm amplitudes for such Fourier modes are needed to instead produce a good linearity curve, meaning a $\gtrsim$50x reduction in phase jitter would be needed to enable higher wavefront order control. See \S\ref{sec: discussion} for further discussion on future solutions to improve this phase jitter limitation.
\subsection{Closed-Loop Operations}
\label{sec: real_time_control}
Building on our measured linear modal basis in \S\ref{sec: linearity}, in this section we will present closed-loop results on-air (\S\ref{sec: on-air}) and with DM-based input AO residual turbulence (\S\ref{sec: turb}).
\subsubsection{On-air Stabilization}
\label{sec: on-air}
\begin{figure}[!h]
    \centering
    \includegraphics[width=0.48\textwidth]{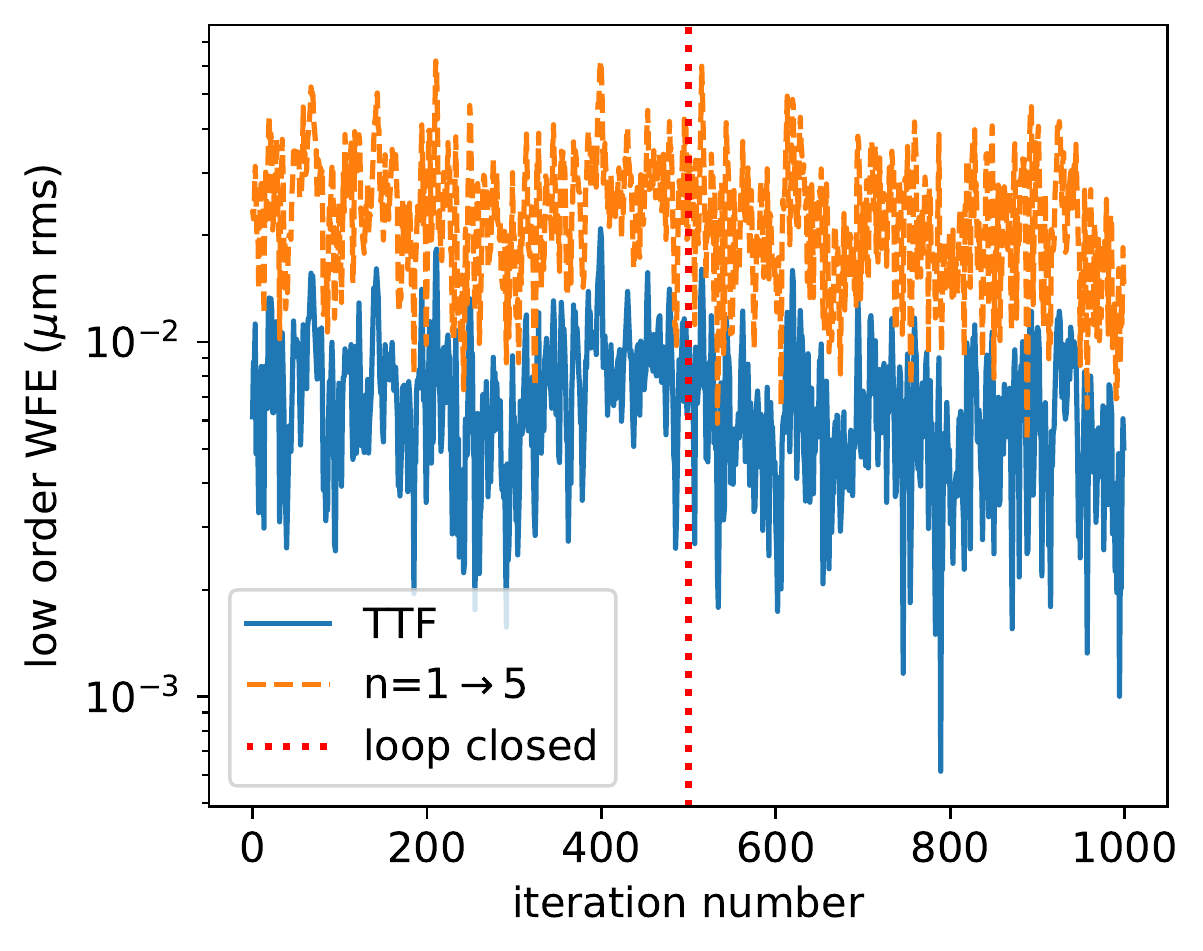}
        \includegraphics[width=0.48\textwidth]{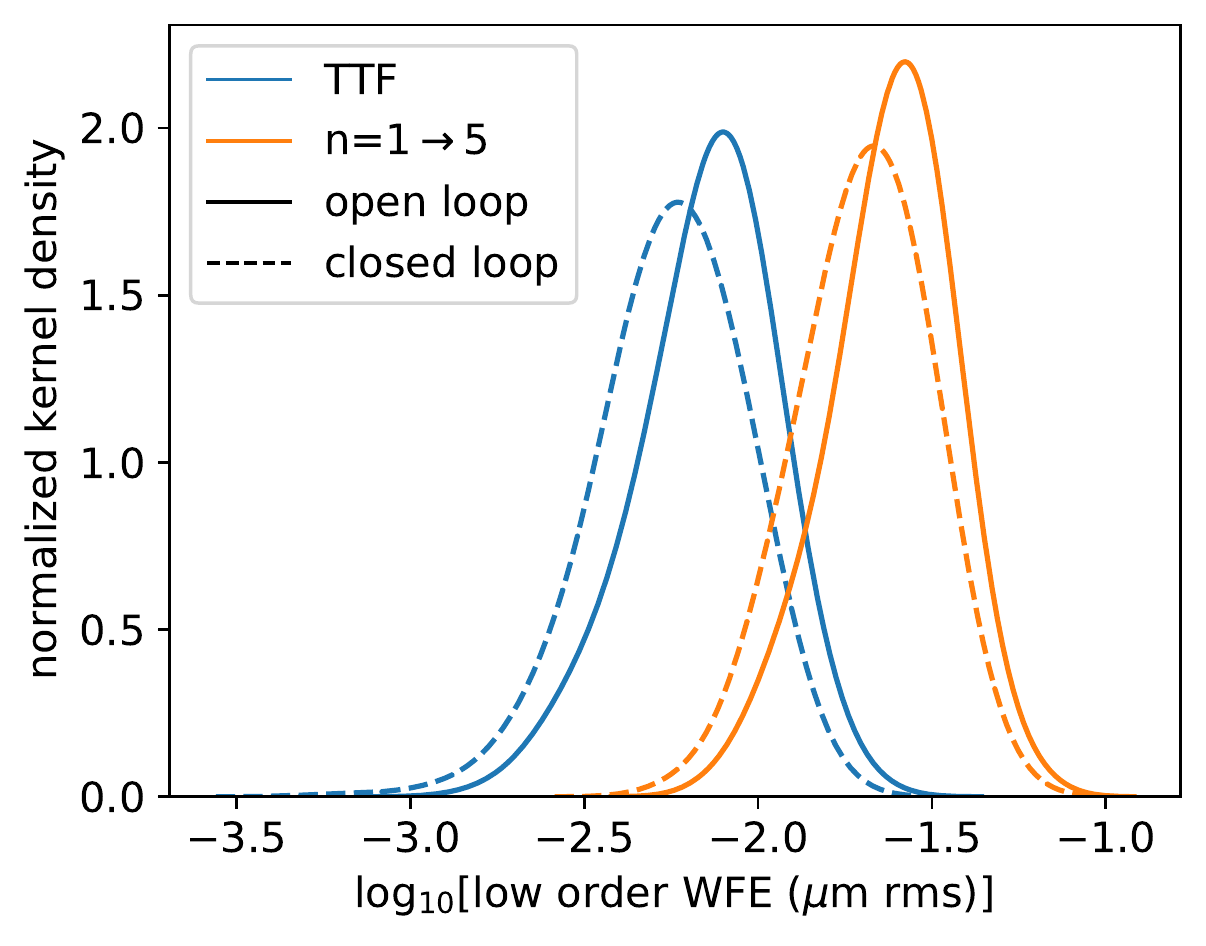}
    \\
    \hspace{-10pt} (a)  \hspace{200pt} (b)
    \caption{Real-time correction of on-air wavefront error (measured by the chopper WFS at a frame rate of about 50 Hz), showing both tip/tilt/focus (TTF) and all controlled modes ($n=1\rightarrow5$), both in the time domain (a) and by WFE distribution (computed using the \texttt{kdeplot} function of the \texttt{seaborn} package; \citealt{seaborn}) comparing the open and closed loop sequences (b; i.e., respectively to the left and right of the dotted line in panel a).}
    \label{fig: on_air}
\end{figure}

On-air closed-loop results are shown in Fig. \ref{fig: on_air}.
Although the stabilized environment from our granite optical table allows only a moderate gain, we still demonstrate a closed-loop WFE reduction of 1.2x for all controlled modes relative to ambient air (i.e., the median of the solid vs. dotted orange distributions in Fig. \ref{fig: on_air}b is 1.2), illustrating the potential of this technique to stabilize quasi-static temporal disturbances on-sky (e.g., thermal variations, flexure, and additional sources of slow beam wander). Closed-loop control was manually tuned for optimal performance using a integrator controller with gains of 0.01, 0.003, and 0.008 for the TTF, low order Zernike, and high order Zernike modal groups, respectively; such low gains were necessary for loop stability due to SEAL's highly stabilized $\sim$nm-level on-air WFE \citep{gerard22}, which as shown is mostly below the noise floor of individual frames of Fig. \ref{fig: on_air}. Note these results help address concerns  that the spinning chopper wheel would be generating additional turbulence, since telemetry from closing the loop on-air clearly shows improvement compared to open-loop.

It is important to note here that we obtained the results in Fig. \ref{fig: on_air} with a high-speed referencing setup designed to minimize chopper blade phase jitter limitations, as discussed in \S\ref{sec: phase_jitter}, and/or integrated effects of on-air turbulence that the chopper may be generating. Specifically, both chop\{$\phi_\text{ref}$\} from equation \ref{eq: chop_im_onsky} and all command matrices are obtained just before the real-time open and closed loop telemetry sequence begins. This strategy ensures that chop\{$\phi_\text{ref}$\} and the command matrices do not suffer from the loss of information due to wavefront ``blurring'' effects, such as chopper blade phase jitter, on-air turbulence evolving, and/or quasi-static evolution of optics, since we found that a reference flat frame and 18 Zernike mode probes (each of which are taken relative to a reference flat before the next mode is probed) acquired at $\sim$50 Hz sufficiently freezes the on-air turbulence and chopper blade phase jitter effects. Other strategies we tried to obtain chop\{$\phi_\text{ref}$\} did not work as well, such as longer exposure acquisition in attempt to ``average out'' these turbulent on-air effects, which, corroborated by Fig. \ref{fig: phase_jitter}, did not work as well. The disadvantage of this fast referencing technique, however, is that single frames can be noisier than other approaches, preventing the deepest convergent WFE this approach can provide; we discuss this further in \S\ref{sec: discussion}.
\subsubsection{Real-time AO Residual Turbulence}
\label{sec: turb}
\begin{figure}[!h]
    \centering
    \includegraphics[width=0.48\textwidth]{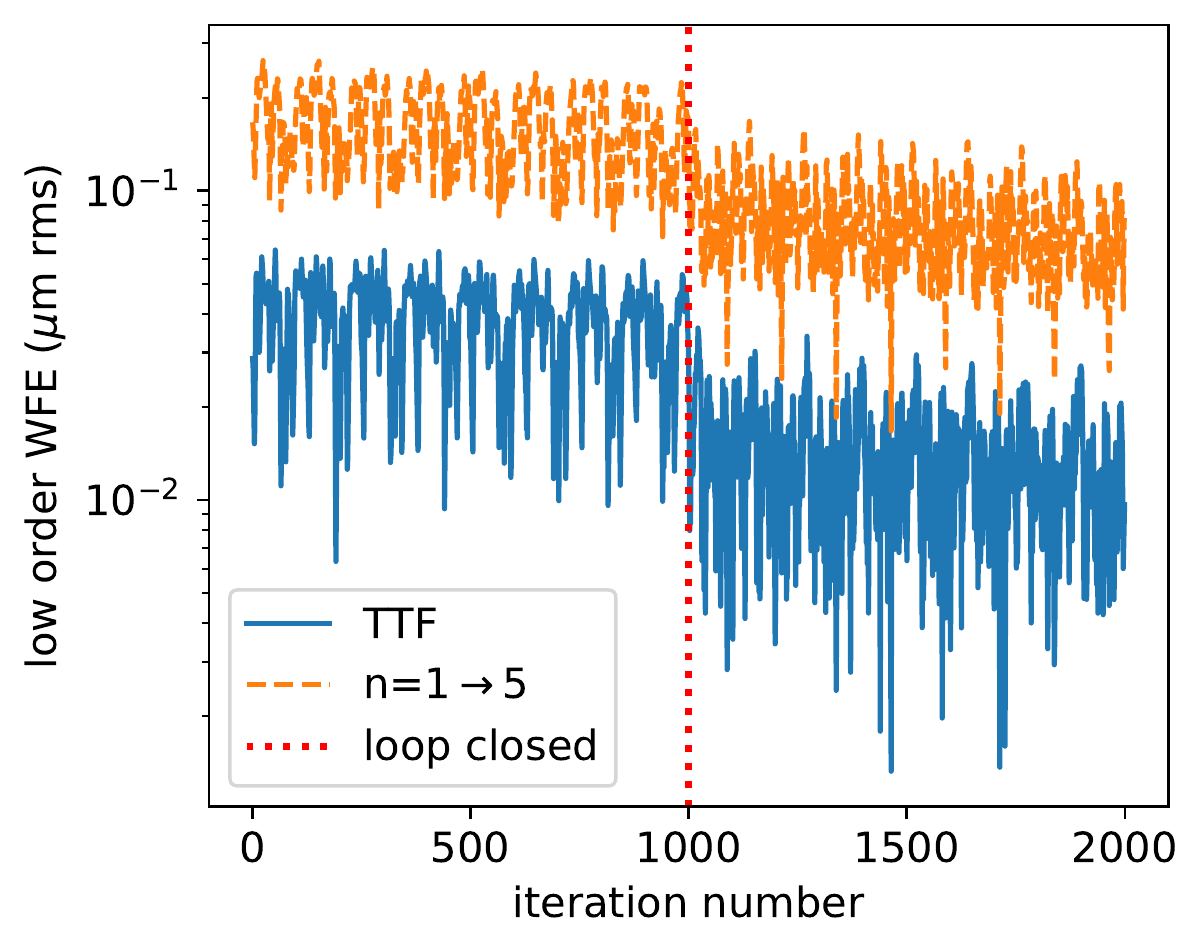}
    \includegraphics[width=0.48\textwidth]{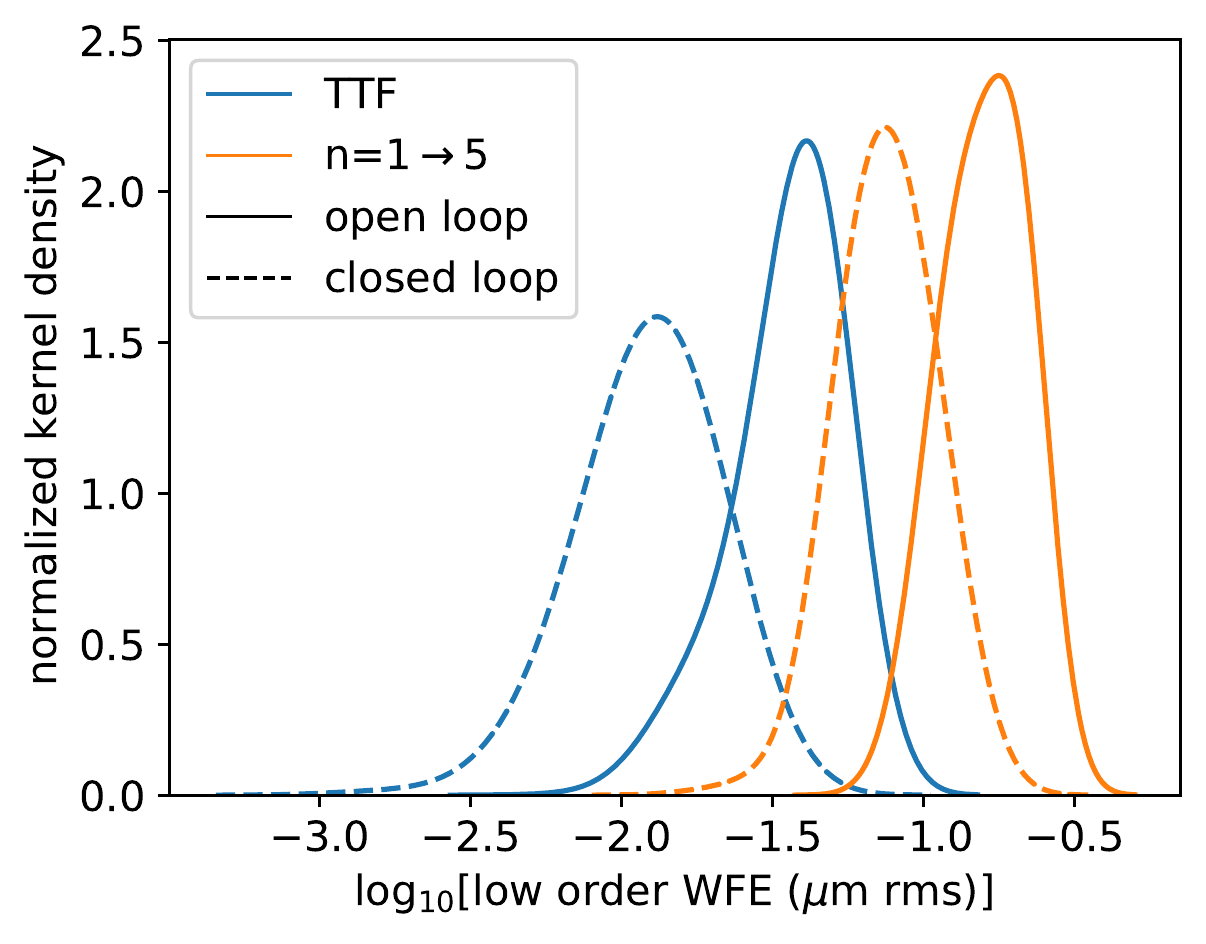}
    \\
    \hspace{-10pt} (a)  \hspace{200pt} (b)
    \\
    \includegraphics[width=1.0\textwidth]{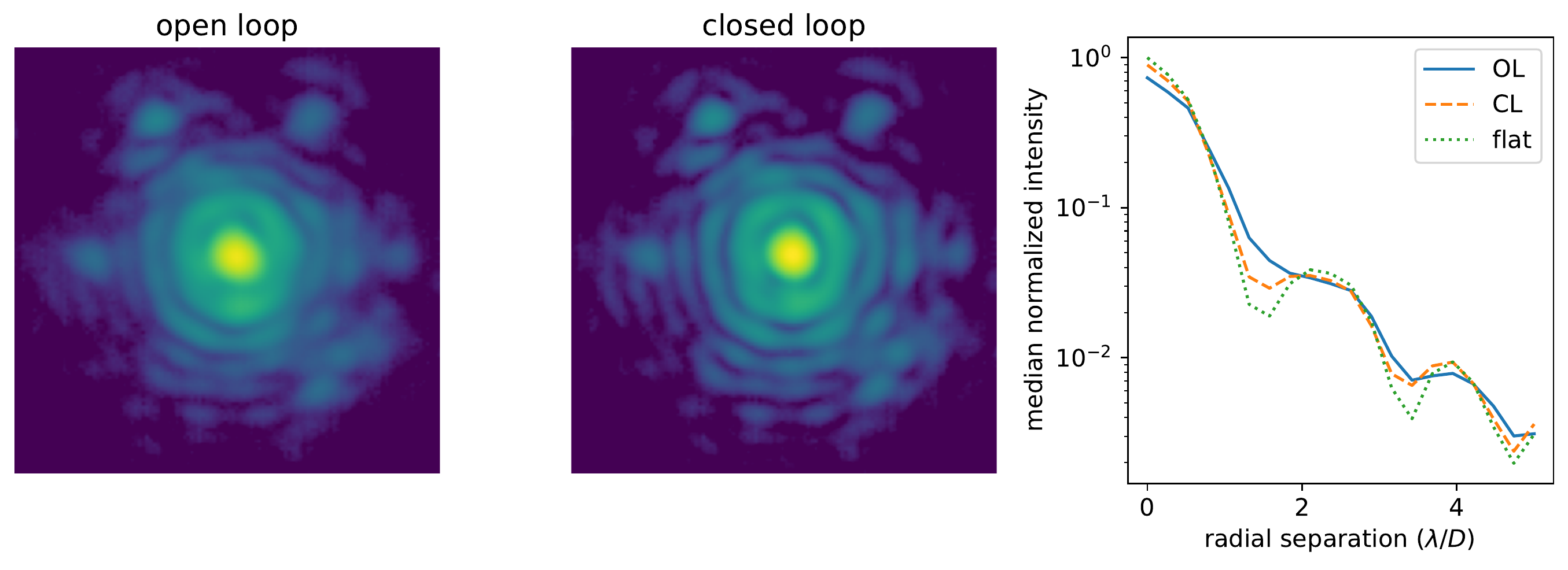}
    \\
    \hspace{-10pt} (c)    
    \caption{Analogous to Figure \ref{fig: on_air}, (a) and (b) show WFE vs. time and open and closed loop distributions for AO residual turbulence applied on the DM. Note that open loop WFE values for the $n=1\rightarrow5$ group may be under- or over-estimated as these levels approach the linear ranges of some modes in Fig. \ref{fig: linearity}. However, panel (c) shows the stack of un-chopped images over the open and closed loop sequences and the corresponding intensity profile for both sequences compared to a static best flat image with no turbulence applied, showing a $\sim$20\% Strehl improvement by closing the loop on AO residuals.}
    \label{fig: on_turb}
\end{figure}
Fig. \ref{fig: on_turb} summarizes our results for closing the loop on DM-induced AO residual turbulence, clearly demonstrating a performance improvement in both WFS telemetry and observed Strehl ratio. Input turbulence is normalized to a 100 nm rms, -2 power law phase screen and translated assuming a single 10 m/s frozen flow ground layer for a 10m telescope, with chopper pair images acquired at 50 Hz. We also use the same high-speed referencing and calibration technique for simulated AO residual turbulence here as described in \S\ref{sec: on-air} for on-air real-time control. Closed-loop control was manually tuned for optimal performance using a leaky integrator controller with gains of 0.7, 0.5, and 0.3 for the TTF, low order Zernike, and high order Zernike modal groups, respectively, and with a leak of 0.95 for all modal groups. 

Fig. \ref{fig: on_turb}a and b shows a total WFE reduction of $\sim$2.2x, roughly consistent with the measured Strehl ratio enhancement from 73 to 92\% (in Fig. \ref{fig: on_turb}c) via the Mar\'{e}cehal approximation. A clear reduction of all controlled modes empirically demonstrates sufficiently minimal cross-talk between different modal groups that use different IM amplitudes and SVD cutoffs, as initially motivated in \S\ref{sec: linearity}. Although it could still be the case that non-linearities dominate the closed-loop error budget in Fig. \ref{fig: on_turb}, such an error budget analysis is beyond the scope of this first introductory paper of the pupil chopping concept, and regardless our results show that these non-linearities are at least $\sim$2.2x below an extreme AO residual WFE, which is an important result. Although the closed loop vs. open loop WFE histogram mean values clearly decrease, their distribution widths remain unchanged, suggesting either our chosen control parameters are not yet fully optimized for robust closed-loop stability and/or similar sources of noise (e.g., chopper phase jitter) account for the distribution width in both cases.

Although these results assume use of a second stage ``cascade'' AO system (\citealt{cascadeAO}; i.e., a separate DM and WFS have produced the residual AO phases used as input into these experiments, this is an important first step in  demonstrating the potential of this technique. The clear benefit in Fig. \ref{fig: on_turb} of closed loop control in the stack of un-chopped images (i.e., science frames) and corresponding radial profiles already illustrates how our existing lab setup could be deployed as a second stage AO system to enhance performance, albeit with AO control limitations as outlined in \cite{cascadeAO}. It is also possible to develop an RTC for two common path WFSs (e.g, a SHWFS and chopper focal plane WFS) to control one or more common path DM(s), e.g., using the framework developed in \cite{gerard21}. This multi-WFS architecture is particularly interesting to explore further for our chopper-based focal plane WFS proposed here, as temporal modulation from the chopper wheel is non-common path to a first-stage WFS (e.g., unlike DM-based phase diversity approaches that would use a common path DM, which further decrease science duty cycle).
\section{Discussion and Future Work}
\label{sec: discussion}
Expanding further on the chopper phase jitter limitations presented in \S\ref{sec: phase_jitter}, the first option to improving chopper performance would be to reduce the phase jitter by at least 50x. Our inquiries to vendors such as Thorlabs and Scitec Instruments indicate that off-the-self optical choppers with such capabilities are not immediately available. More custom options have nonetheless demonstrated the ability to reach such limits \citep{johnson22} and should be explored further. However, the DM-based chopping approach presented in \cite{gerard22spie} does not suffer from a similar phase jitter problem. Of particular importance to highlight after these demonstrations is the potential for non-coronagraphic focal plane WFS applications, including single conjugate AO, laser guide star (LGS) AO (compatible with off-axis sensing), wavefront sensing in crowded fields, and tomographic reconstruction. All of these applications are enabled simply by recording consecutive chopped and un-chopped focal plane images, with the latter used for science. LGS and off-axis sensing applications would additionally require the detector to image the off-axis guide star (and for LGSs placed to the appropriate position to image the focal plane), Crowded field sensing would require point-source detection algorithms to accommodate a different position distributions of stars for a given on-sky pointing and/or for LGS tomography pre-defined image positions. Multiple sources in crowded fields could also potentially limit the number of controllable modes, as in principle the same pixels on a detector cannot be used for wavefront reconstruction from two different incoherent sources without a multi-star wavefront control strategy \citep{sirbu2017}. Tomographic reconstruction with such multiple sources could then enable wide-field AO correction, either for ground-layer AO for a single pupil-conjugated DM or for multi-congugate AO. Relatedly, guiding on resolved astrophysical objects should be further explored.

There are many areas to further explore and develop such that this pupil chopping technique can be operational on-sky. Perhaps the most pressing is overcoming the $\sim$10 c/p sensing limit presented in \S\ref{sec: ho_sim}, currently limiting higher order wavefront control to DMs with less than $20\times20$ actuators. For the DM-based chopper technique, it is possible that more complex chopper ``ridge-line''\footnote{By ``ridge-line,'' we mean the shape of the line defining the transition between the modulated and un-modulated pupil fractions. In this paper we have just considered this to be a straight line.} geometries could enable improved high order measurement sensitivity. Although this would be a challenging endavor for an optical chopper blade, it is a simple application for a DM, and particularly for a segmented DM. Expanding on the discussion in \S\ref{sec: on-air}, generating chop\{$\phi_\text{ref}$\} from simulation and/or using a fully synthetic interaction matrix would be interesting to explore further. Like the fast referencing approach previously discussed, this approach does not suffer from wavefront blurring effects, but furthermore it is not photon and/or detector noise-limited, potentially allowing deeper achievable closed-loop convergent WFEs but with the added risk of additional model-based errors. Such model errors could be less problematic for this method as presented in this manuscript compared to coronagraphic focal plane wavefront sensing techniques that have additional degrees of freedom to model \citep{potier20}. Coronagraphic applications of this technique should also be explored; in principle the chopper temporal modulation concept can remain identical to what we presented in this paper, but additional coronagraph-specific questions should be investigated (e.g., chopping in an apodizer vs. Lyot stop plane, and WFS linearity and sensitivity dependence on different coronagraph designs). Higher duty cycle chopper operations could also be explored further; the 2\% duty used in laboratory testing presented in this paper (see \S\ref{sec: setup}) is effectively generating a PSF with an obscured aperture at a single chop fraction, but higher duty cycles would cover a range of fractions over a single camera integration due to the continuous motion of the chopper blade. Although these effects could be simulated and tested, again the DM-based chopping approach introduced in \cite{gerard22spie} does not have this problem and can in principle reach higher than 50 \% duty cycles with a custom camera read out scheme. Another interesting area to further explore is absolute phase retrieval; we have only presented this technique so far as differential relative to a pre-calibrated best flat enabled by some other method, but this method could in principle enable non-linear (e.g., Gerchberg–Saxton) absolute wavefront reconstruction   (either with a single chopper pair for the chopper blade ``amplitude diversity'' approach and/or a single chopper image for the DM-based chopper ``phase diversity'' approach). Performance over a large spectral bandpass should be explored; although PSF images inherently suffer from chromatic magnification without a Wynne corrector, in this technique the wavefront diversity is applied in the pupil plane, potentially enabling robustness to higher spectral bandwidth operations with this technique.
\section{Conclusion}
\label{sec: conclusion}
We have presented a new focal plane wavefront sensing technique, which uses an optical chopper designed to partially block and then unblock the pupil in consecutive frames while recording synchronized focal plane images (\S\ref{sec: concept}). This pupil chopping provides sufficient amplitude diversity to reconstruct wavefront modes $\lesssim$10 c/p (\S\ref{sec: ho_sim}). It is also optimal for high speed wavefront control, requiring only two consecutive images (one of which is the science image) and a MVM to generate DM commands for residual AO correction (\S\ref{sec: lowfs_recon_sim}). In addition to simulations of this new concept, we presented laboratory results using SEAL (\S\ref{sec: lab}), summarized as follows:
\begin{enumerate}
\item In \S\ref{sec: linearity} we measured good linearity for reconstructed low order Zernike modes ($n<6$), but higher order Zernikes were not measurable due to chopper phase jitter (\S\ref{sec: phase_jitter}; but see \S\ref{sec: discussion}).
\item We first closed the loop on air at $\sim$50 Hz (i.e., consecutive image acquisition at 100 Hz) in \S\ref{sec: on-air}, clearly improving the WFE and demonstrating the potential for real-time correction of quasi-static errors.
\item We next closed the loop in \S\ref{sec: turb} on DM-induced AO residual turbulence, also at 50 Hz, showing a clear performance gain, including by a measured 20\% Strehl ratio increase in closed vs. open loop cases.
\end{enumerate}
More topics need to be addressed before this technique can be deployed at observatories on-sky (\S\ref{sec: discussion}), but thus far we foresee no showstoppers towards enabling this. The power of this technique---particularly in the DM-based chopping approach presented first in \cite{gerard22spie} and in a forthcoming more-detailed paper (Soto, Gerard et al., in prep)---is its simplicity. Hardware simplicity (only requiring a focal plane imager in addition to a conventional AO system), software simplicity (operating with a linear MVM, in comparison to other iterative, non-linear focal plane WFS techniques), and broad compatibility (with applications non-coronagraphic and/or coronagraphic systems, LGS and/or natural guide star systems, and correction of high-speed atmospheric residuals and/or quasi-static WFEs) illusrate a promising and powerful potential for this new technique.
\section*{Acknowledgements}
We gratefully acknowledge research support of the University of California Observatories for funding this research. Author B. Gerard thanks the 2018 SCExAO team for hosting discussions that led to the pupil chopping concept. This work performed under the auspices of the U.S. Department of Energy by Lawrence Livermore National Laboratory under Contract DE-AC52-07NA27344. The document number is LLNL-JRNL-843401. The authors thank the anonymous reviewer for their detailed consideration and feedback of this manuscript.

\appendix
\section{Linear Least-Squares Reconstructor Summary}
\label{sec: mvm}
Below we summarize the widely-used calibration routine to convert WFS measurememts in real-time into closed-loop DM commands as applied to this paper.
\begin{enumerate}
\item For $k$ DM modes (e.g., Zernike modes), $w$ from equation \ref{eq: chop_im_onsky} is recorded, for low order modes selecting pixel values within a given control radius of the detector focal plane optical axis (e.g., if controlling up to 5 radial orders of Zernike modes, only using $w$ pixel values within a 5 $\lambda/D$ radius from the optical axis). Thus, for DM each mode we generate a vector $w_k$ of dimensions $1\times j$, where $j$, is the number of $w$ pixel values used for a given image.
\item an Interaction Matrix (IM) concatenates $w_k$ for all $k$ modes into a $k\times j$ matrix. A reference vector of DM actuator commands for all modes, $R$, is also saved during this modal calibration, generating a $k\times p$ matrix in units of DM command space (e.g., volts), where $p$ is the number of actuators within the two-dimensional DM pupil footprint. 
\item{ The command matrix (CM), used to convert $w$ space to DM command space assuming linearity between the two, is then
\begin{equation}
\text{CM}= \text{IM}^\dagger \; \cdot  \; - R,
\label{eq: cm}
\end{equation}
where $\cdot$ and $^\dagger$ represent a matrix dot product and a matrix pseudo inverse process such as Singular Value Decomposition (SVD), respectively. 
}
\item{ Open loop DM commands are then reconstructed by an on-sky input $w$ via
\begin{equation}
\text{DMC}_\text{open-loop} = w_\text{on-sky}\; \cdot \; CM
\label{eq: mvm}
\end{equation}
where $w_\text{on-sky}$ is a $1\times j$ vector, representing a real-time $w$ value, and DMC$_\text{on-sky}$ is a $1\times p$ vector of real-time DM commands (DMCs). 
}
\item{ Finally, closed-loop real-time control with a leaky integrator is given by
\begin{equation}
\text{DMC}_{n}=l\; \text{DMC}_{n-1}\; +\; g\; \text{DMC}_\text{open-loop},
\label{eq: integrator}
\end{equation}
where DMC$_n$ is the current frame of DMCs, DMC$_{n-1}$ is the previous frame of DMCs, and $g$ and $l$ are the integrator controller gain and leak parameters, respectively.
}
\end{enumerate}
\bibliography{refs}
\end{document}